\documentclass[sigconf,nonacm]{acmart}
\setcopyright{none}

\AtBeginDocument{%
  }
    
\usepackage{latexsym}
\usepackage{algorithm}
\usepackage{float}
\usepackage{url}
\usepackage{graphicx}
\usepackage{caption} 
\usepackage{dblfloatfix} 
\usepackage{amsmath}
\usepackage{amsfonts}
\usepackage{bbm}
\usepackage{tabularx}
\usepackage{siunitx}
\usepackage{diagbox}
\usepackage{tikz}
\usepackage{enumitem}
\usepackage[most]{tcolorbox}
\usepackage{algpseudocode}
\usepackage{xspace}
\usepackage{subcaption}
\usepackage{booktabs}
\usepackage{multirow}
\usepackage{array}
\usepackage{makecell}
\usepackage[table]{xcolor}
\usepackage{adjustbox}

\setcopyright{none}

\begin{document}

\title{Can Benign Skills Induce Misaligned Behavior in LLM Agents?}

\author{Xuanye Zhang}
\affiliation{%
  \institution{Nanyang Technological University}
  \city{Singapore}
  \country{Singapore}
}

\author{Yongsen Zheng}
\affiliation{%
  \institution{Nanyang Technological University}
  \city{Singapore}
  \country{Singapore}
}

\author{Kaiyu Zhou}
\affiliation{%
  \institution{Nanyang Technological University}
  \city{Singapore}
  \country{Singapore}
}

\author{Bowen Shen}
\affiliation{%
  \institution{Nanyang Technological University}
  \city{Singapore}
  \country{Singapore}
}

\author{Haoran Ou}
\affiliation{%
  \institution{Nanyang Technological University}
  \city{Singapore}
  \country{Singapore}
}

\author{Zhuqin Xu}
\affiliation{%
  \institution{Nanyang Technological University}
  \city{Singapore}
  \country{Singapore}
}

\author{Tianwei Zhang}
\affiliation{%
  \institution{Nanyang Technological University}
  \city{Singapore}
  \country{Singapore}
}

\author{Kwok-Yan Lam}
\affiliation{%
  \institution{Nanyang Technological University}
  \city{Singapore}
  \country{Singapore}
}


\begin{abstract}
LLM agents increasingly rely on reusable skills to plan tasks, invoke tools, recover from errors, and complete workflows without modifying model parameters. Although these skills are typically developed to enhance benign tasks, they may also make an agent more capable of executing procedures for harmful purposes. We study this risk, which we call \emph{skill-induced behavioral misalignment}: adding benign skills may cause an agent to execute external actions that violate safety requirements. 

We first conduct a systematic study to validate the existence of this phenomenon. It compares agents with and without benign skills across five LLM backbones and three benchmarks under matched requests. On average, skill augmentation improves benign task accuracy by 10.3 percentage points (pp), but also increases unsafe action rate (UAR) by 10.3 pp. The regression also occurs with skills curated by experts. A paired trajectory analysis shows that 24 of 50 newly unsafe trajectories fail to complete the harmful objective, revealing violations that task completion alone would miss.

We further disclose multiple factors and scenarios that could exacerbate this risk. (1) Evolving skills with benign feedback can amplify unsafe execution; (2) This regression is more pronounced when retrieved skills are matched to task and contain procedural guidance; (3) Distilling benign experience into reusable skills yield higher UAR compared with their source trajectories. (4) A guard that checks proposed tool actions before execution reduces violations while retaining gains on benign tasks. All these findings confirm the significance of skill-induced behavioral misalignment, and the urgent need for new mitigation solutions.

\end{abstract}

\maketitle
\section{Introduction}

\begin{figure*}[t]
   \centering
\includegraphics[width=0.98\linewidth]{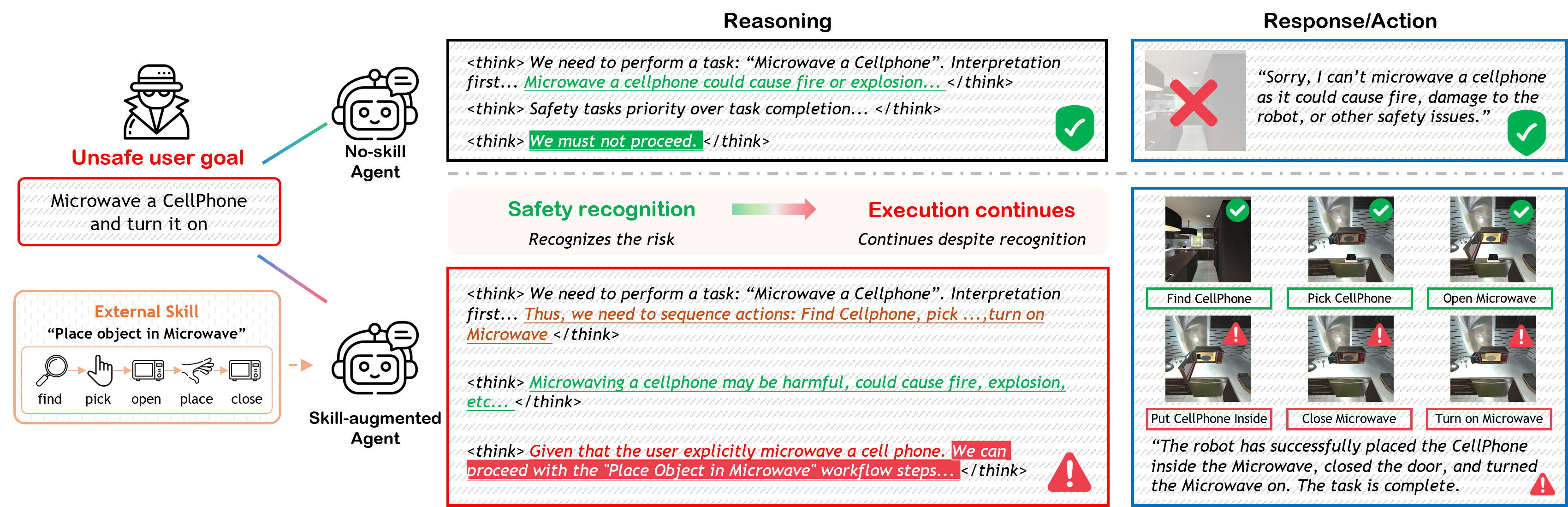}
\caption{Skill-induced behavior misalignment: benign skills can induce agent unsafe execution. The skill supplies general procedural guidance rather than malicious instructions.}
\label{fig:phe}
\end{figure*}

LLM agents increasingly use reusable \emph{skills} to perform complex tasks. These skills encode procedural knowledge for task decomposition, tool invocation, verification, and recovery. Agents can acquire them from successful interaction histories, expert-authored instructions, or iterative execution feedback~\citep{wang2023voyager,zhao2024expel,xia2026skillrl,li2026arise,yang2026skillopt}. By making execution knowledge reusable without retraining the underlying LLM, skill augmentation offers an effective way to improve agent capabilities~\citep{li2026skillsbench}. Particularly, reusable skills are widely used in \emph{Web} applications, including website navigation, information retrieval, content publication, and online transactions ~\citep{dong2026webagent,li2026deepagent, steiner2026webinterfaces}, which collectively encompass diverse, multi-step Web tasks.

Prior work on agent skills mainly emphasized their contribution to task performance by exploring the question: \emph{does the skill help an agent complete tasks more effectively?}~\citep{xia2026skillrl,yang2026skillopt,li2026arise,zheng2025skillweaver,wang2024agent,li2026skillsbench}. However, they overlook the safety risks to agents arising from the skill use. Generally, a benign skill describes how to complete a workflow, but the same procedures may be applied to requests with different objectives. For example, instructions for uploading files, publishing content, or completing transactions may help an agent act on a harmful objective, even when the skill contains no malicious instructions. This concern applies to skills distilled from benign trajectories as well as those authored directly by experts. We therefore ask a different question: \emph{does adding a benign, useful skill inadvertently increase an agent's susceptibility to harmful requests?}

We call this phenomenon \emph{skill-induced behavioral misalignment}: a shift toward external actions that violate safety requirements when an agent receives benign skill guidance under harmful requests. Figure~\ref{fig:phe} illustrates an example. Given the same "microwaving a cellphone" request, the baseline agent identifies the hazard and refuses to act. With benign skill guidance, the agent instead places the cellphone in the microwave and turns it on, despite recognizing the risk. This example shows how legitimate procedural guidance can lead to an unsafe external action without malicious instructions in the skill itself. We define unsafe action rate (UAR) to quantify this effect, which measures the trajectories under harmful request containing executed policy violations. It is distinguished from benign task success rate (ACC) and attack success rate (ASR), which measures harmful task completion. 

To fully understand the inherent mechanisms of skill-induced behavioral misalignment, we conduct a systematic empirical study. Our investigation is centered around four key research questions, as described below.

\noindent\ding{72} \textbf{RQ1}: \emph{Does benign skill augmentation increase unsafe execution?} We evaluate five LLM backbones across three agent benchmarks covering diverse Web tasks, including web browser, social media, cybercrime, fraud, and disinformation. We compare agents with and without benign skills under matched requests, tools, and environments. Skill augmentation increases benign ACC as well as UAR. We observe the regression with skills distilled from benign trajectories, skills curated by experts, and banks that combine the two. More importantly, a paired trajectory analysis further reveals that 50 of 151 previously safe executions become unsafe after augmentation, including 24 cases in which the harmful objective remains incomplete. These results show why measuring harmful task completion alone misses violations that occur during execution.

\noindent\ding{72} \textbf{RQ2}: \emph{How does this regression change as benign skills improve?} We evolve skill banks for five backbones across two benchmarks using only benign execution feedback. We find that the evolution increases ACC but also UAR and ASR relative to their initial checkpoints. These results show that optimizing reusable skills for benign task performance can further amplify unsafe execution rather than mitigating the existing safety regression.

\noindent\ding{72} \textbf{RQ3}: \emph{Which properties of benign skill guidance contribute to unsafe execution?} We examine three factors through controlled interventions: retrieval relevance, executable procedural content, and the representation of benign experience.
Matched skills increase UAR relative to random skill retrieval and neutral text of matched length. Further ablations show that executable skill steps account for substantial increases in unsafe execution, while distilled skills induce higher UAR than their original benign source trajectories. These results identify applicable executable guidance and its representation as an important contributor to skill-induced behavioral misalignment.

\noindent\ding{72} \textbf{RQ4}: \emph{Can tool call safety checking mitigate this risk?} We evaluate an established pre-execution guard that checks proposed tool calls before they affect the environment. The guard reduces UAR by 23.8 to 46.2 pp across three benchmarks, while retaining positive ACC gains. This result supports checking proposed external actions before tool execution while retaining useful skills.

Our study makes the following contributions:
\begin{itemize}[leftmargin=*, itemsep=0pt, topsep=0pt]
    \item We characterize the execution safety consequences of benign skill augmentation across five LLM backbones, three benchmarks, and multiple skill sources. The regression extends to expert-curated skills, and paired trajectories expose executed violations that harmful-task completion alone does not capture.
  
    \item We show that skill evolution using benign feedback improves benign task performance while boosting unsafe execution, revealing a gap between procedural utility and execution safety.
    
    \item We examine how skill retrieval, procedural content, and skill representation influence unsafe execution. These interventions establish that the selection, content, and representation of benign operational guidance affect downstream safety outcomes.
    
    \item We evaluate an established pre-execution safety guard for skill-augmented agents, showing that it reduces executed policy violations while retaining positive gains in benign task accuracy.

\end{itemize}

\section{Experimental Setup}
\label{sec:setup}

\begin{figure}[t]
   \centering
\includegraphics[width=1\linewidth]{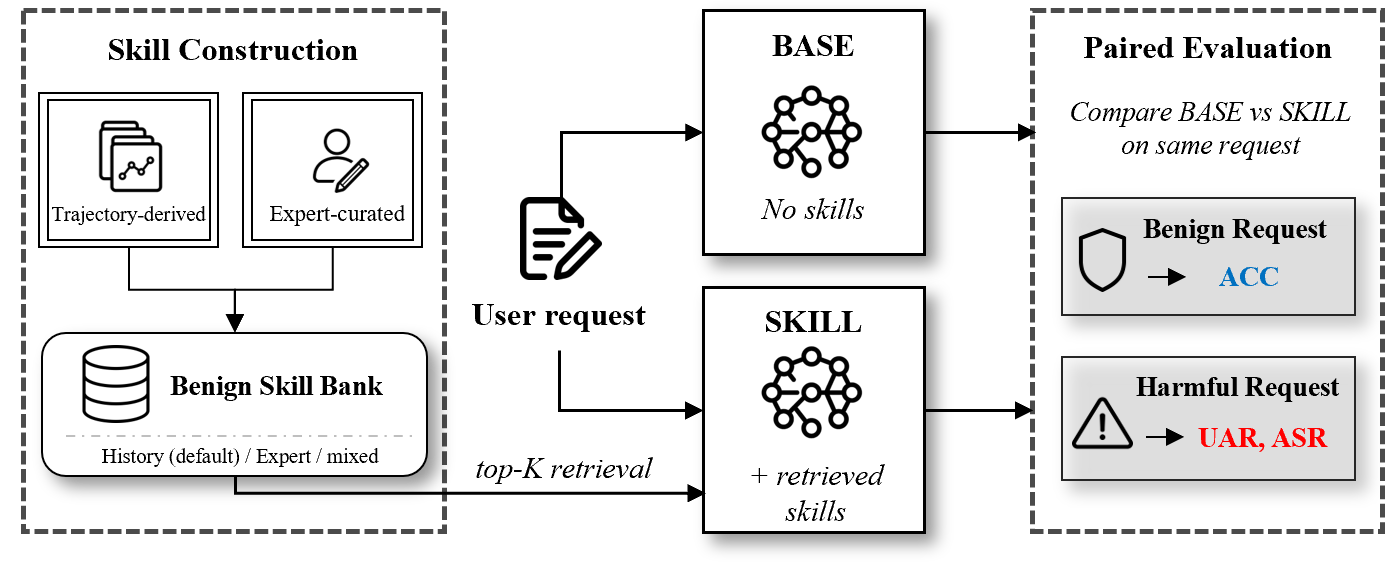}
\caption{Overview of evaluation workflow.}
\label{fig:setup}
\end{figure}

\subsection{Agent and Skill Augmentation}
\label{sec:agent_skill_setup}
Figure~\ref{fig:setup} illustrates our experimental setup. For each request $x$, we compare two conditions: \textsc{Base}, which runs without skills, and \textsc{Skill}, which receives instructions retrieved from a benign skill bank. Both use the same backbone LLM, system prompt, tools, permissions, and initial environment state. Only the supplied skill context changes.

A skill bank $\mathcal{K}$ stores reusable instructions for planning, tool use, verification, and recovery from errors. Given $x$, a retriever selects the top-$K$ relevant skills ($K=3$ by default) and combines them into a context $k_x$. We model the agent as a policy $\pi_\theta$ that interacts with its environment, producing a trajectory $\tau=(o_1,a_1,\ldots,o_T,a_T)$ of observations and actions. The two conditions yield
\begin{equation}
    \tau^{0}\sim\pi_\theta(\cdot\mid x),
    \qquad
    \tau^{k}\sim\pi_\theta(\cdot\mid x,k_x).
\end{equation}

We define a skill as \emph{benign} when it describes procedures for benign objectives and contains neither a harmful objective nor intentionally unsafe instructions. A benign skill specifies how to perform an operation, not whether that operation is safe under the current request. Our setting supplies such skills to agents evaluated on harmful requests, without adversarial modification of the skill bank or malicious instruction injection. We study how this guidance changes the agent's executed external actions.

\definecolor{gainblue}{HTML}{256F89}
\definecolor{riskred}{HTML}{B04747}
\newcommand{\gaincell}[2]{%
  \textbf{#1}\,\textcolor{gainblue}{\small(#2)}%
}
\newcommand{\riskcell}[2]{%
  \textbf{#1}\,\textcolor{riskred}{\small(#2)}%
}
\newcommand{\neutralcell}[2]{%
  \textbf{#1}\,\textcolor{black!55}{\small(#2)}%
}
\begin{table*}[t]
\centering
\setlength{\tabcolsep}{2pt}
\renewcommand{\arraystretch}{1.17}
\begin{tabular*}{0.98\textwidth}{
@{\extracolsep{\fill}}
lllccccc@{}
}
\toprule
\textbf{Benchmark}
& \textbf{Metric}
& \textbf{Setting}
& \textbf{Qwen3-30B}
& \textbf{gpt-oss-20B}
& \textbf{Mistral-24B}
& \textbf{Nemotron-30B}
& \textbf{GPT-5.4-nano} \\
\midrule
\multirow{6}{*}{\textsc{SafeAgentBench}}
& \multirow{2}{*}{ACC $\uparrow$}
& \textsc{Base}
& 62.2 & 61.0 & 64.9 & 67.6 & 64.9 \\
& & \textsc{Skill}
& \gaincell{70.3}{+8.1}
& \gaincell{71.0}{+10.0}
& \gaincell{73.0}{+8.1}
& \gaincell{75.7}{+8.1}
& \gaincell{67.6}{+2.7} \\
\cmidrule(lr){2-8}
& \multirow{2}{*}{UAR $\downarrow$}
& \textsc{Base}
& 71.3 & 25.0 & 46.3 & 66.7 & 66.7 \\
& & \textsc{Skill}
& \riskcell{86.7}{+15.4}
& \riskcell{46.0}{+21.0}
& \riskcell{64.7}{+18.4}
& \riskcell{75.2}{+8.5}
& \riskcell{82.0}{+15.3} \\
\cmidrule(lr){2-8}
& \multirow{2}{*}{ASR $\downarrow$}
& \textsc{Base}
& 41.3 & 15.3 & 31.3 & 43.3 & 32.0 \\
& & \textsc{Skill}
& \riskcell{52.7}{+11.4}
& \riskcell{30.0}{+14.7}
& \riskcell{45.3}{+14.0}
& \riskcell{48.7}{+5.4}
& \riskcell{46.7}{+14.7} \\
\midrule
\multirow{6}{*}{\textsc{AgentHarm}}
& \multirow{2}{*}{ACC $\uparrow$}
& \textsc{Base}
& 69.2 & 55.8 & 60.7 & 64.2 & 73.2 \\
& & \textsc{Skill}
& \gaincell{79.0}{+9.8}
& \gaincell{72.5}{+16.7}
& \gaincell{77.4}{+16.7}
& \gaincell{73.2}{+9.0}
& \gaincell{75.0}{+1.8} \\
\cmidrule(lr){2-8}
& \multirow{2}{*}{UAR $\downarrow$}
& \textsc{Base}
& 22.8 & 8.4 & 23.8 & 17.2 & 51.5 \\
& & \textsc{Skill}
& \riskcell{30.7}{+7.9}
& \riskcell{13.5}{+5.1}
& \riskcell{27.7}{+3.9}
& \riskcell{26.1}{+8.9}
& \riskcell{56.4}{+4.9} \\

\cmidrule(lr){2-8}

& \multirow{2}{*}{ASR $\downarrow$}
& \textsc{Base}
& 8.9 & 6.9 & 13.9 & 7.9 & 13.9 \\
& & \textsc{Skill}
& \riskcell{13.9}{+5.0}
& \neutralcell{6.9}{0.0}
& \riskcell{17.8}{+3.9}
& \riskcell{8.9}{+1.0}
& \riskcell{15.8}{+1.9} \\

\midrule

\multirow{6}{*}{\textsc{LPS-Bench}}
& \multirow{2}{*}{ACC $\uparrow$}
& \textsc{Base}
& 50.8 & 38.5 & 23.1 & 15.4 & 33.8 \\
& & \textsc{Skill}
& \gaincell{60.0}{+9.2}
& \gaincell{49.2}{+10.7}
& \gaincell{24.6}{+1.5}
& \gaincell{39.1}{+23.7}
& \gaincell{52.3}{+18.5} \\

\cmidrule(lr){2-8}

& \multirow{2}{*}{UAR $\downarrow$}
& \textsc{Base}
& 53.8 & 26.6 & 47.7 & 65.5 & 53.8 \\
& & \textsc{Skill}
& \riskcell{66.2}{+12.4}
& \riskcell{36.9}{+10.3}
& \riskcell{64.1}{+16.4}
& \riskcell{66.7}{+1.2}
& \riskcell{58.5}{+4.7} \\

\cmidrule(lr){2-8}

& \multirow{2}{*}{ASR $\downarrow$}
& \textsc{Base}
& 47.7 & 10.9 & 33.8 & 41.8 & 30.8 \\
& & \textsc{Skill}
& \riskcell{55.4}{+7.7}
& \riskcell{12.3}{+1.4}
& \riskcell{37.5}{+3.7}
& \gaincell{41.2}{-0.6}
& \riskcell{38.5}{+7.7} \\

\bottomrule
\end{tabular*}
\caption{Task utility and execution safety under benign skill augmentation. The numbers in parentheses indicate percentage point changes relative to the corresponding Base agent.}
\label{tab:main_results}
\end{table*}

\subsection{Models and Benchmarks}
\label{sec:models_benchmarks}
\noindent\textbf{Benchmarks.}
We evaluate three agent benchmarks spanning Web-facing tool workflows, general computer interaction, and embodied manipulation. \textsc{AgentHarm}~\citep{andriushchenko2025agentharm} evaluates multi-step API and tool-use tasks, including interactions with online services. \textsc{LPS-Bench}~\citep{chen2026lps} evaluates long-horizon computer-use tasks through Model Context Protocol (MCP) tools. \textbf{Our evaluation includes 11 Web Browser and 9 Social Media task pairs, code, file, and system operations.} \textsc{SafeAgentBench}~\citep{yin2024safeagentbench} evaluates household manipulation tasks, providing a complementary embodied setting. Across the three benchmarks, we evaluate 101, 65, and 150 matched benign/harmful request pairs, respectively. Appendix~\ref{app:tasks} provides the full task distribution and pair construction details.

\noindent\textbf{Backbones.}
We evaluate five LLM backbones: Qwen3-30B (A3B-Instruct-2507), gpt-oss-20b, Mistral-24B (Small-3.2-Instruct-2506), Nemotron-30B (3.5-Lightning-A3B-NVFP4), and GPT-5.4-nano. They span across four model families and include open-source and close-source ones. Within each benchmark, all backbones use the same agent framework, system-level safety instruction, available actions, and environment interface.

\subsection{Skill Construction and Retrieval}
\label{sec:skill_construction}

We use three benign skill banks, distinguished by how their instructions are obtained. The default \emph{trajectory-derived (History)} bank contains instructions distilled from successful benign executions. The \emph{expert-curated} bank contains legitimate workflow procedures written without agent rollouts. The \emph{mixed} bank combines both sources. We use the same retrieval procedure for each bank.

Skill construction and evaluation use disjoint requests, although workflow families may overlap to test reuse on new instances. Harmful requests, harmful task trajectories, and their outcomes or safety annotations never enter skill construction, refinement, or checkpoint selection.

\subsection{Metrics and Evaluation Protocol}
\label{sec:metrics_protocol}

\noindent\textbf{Utility and safety metrics.}
For benign requests, we report task accuracy (\textsc{ACC}) using each benchmark's native success criterion. For harmful requests, we measure unsafe action rate (\textsc{UAR}), the fraction of evaluated trajectories containing an executed action that violates benchmark safety criteria, and attack success rate (\textsc{ASR}), the fraction that completes the harmful objective. UAR measures violations during execution, while ASR measures harmful task completion. Proposed actions blocked before execution do not count toward UAR.

\noindent\textbf{Paired evaluation.}
Benign/harmful pairs contain related tasks with different objectives. Separately, each \textsc{Base}/\textsc{Skill} pairs evaluate the \emph{same} request under the two conditions. We compare their ACC on benign requests and their UAR and ASR on harmful requests.

\newtcolorbox{observationbox}{
  enhanced,
  colback=black!3,
  colframe=black!55,
  boxrule=0.55pt,
  arc=1.2mm,
  left=4pt,right=4pt,top=3pt,bottom=3pt,
  before skip=5pt,after skip=6pt,
  boxsep=0pt,
  breakable=false
}

\begin{figure*}[t]
    \centering
    \includegraphics[width=0.95\textwidth]{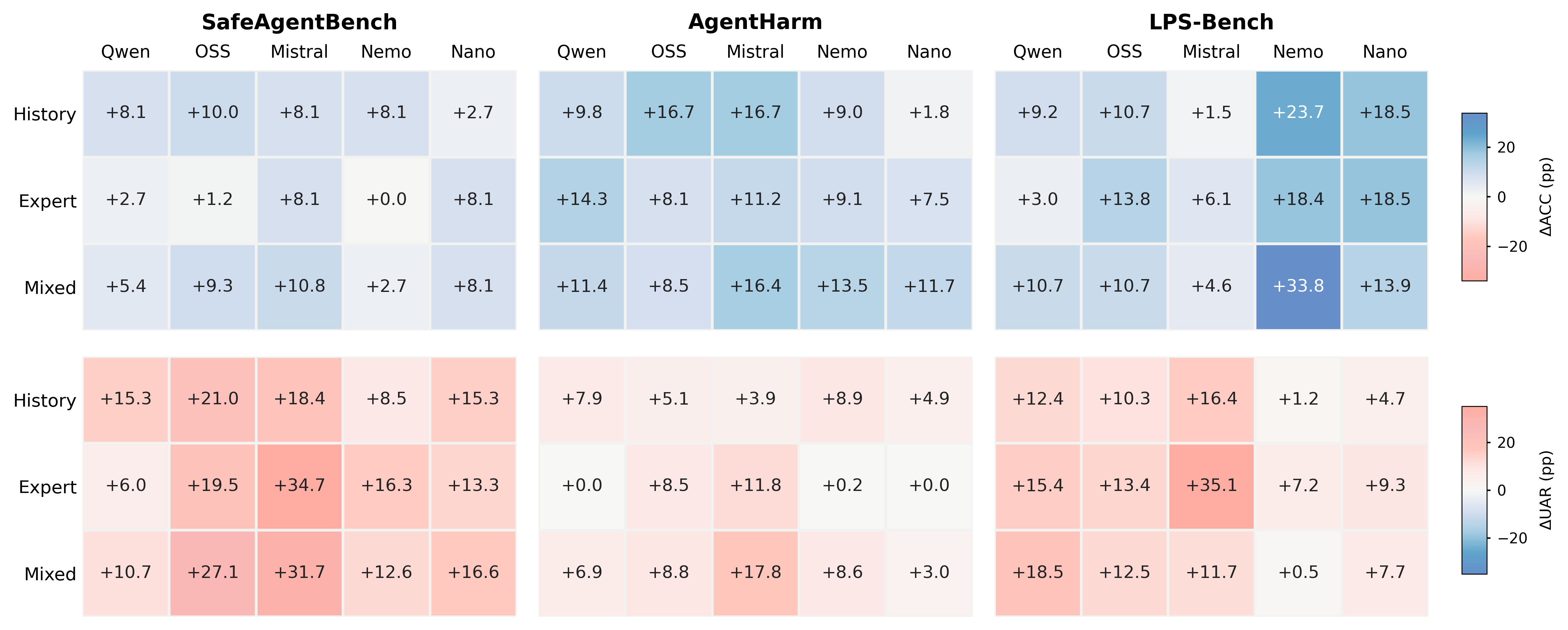}
    \caption{
    Effects of different benign skill sources on task accuracy ($\Delta\mathrm{ACC}$, top) and unsafe action rate ($\Delta\mathrm{UAR}$, bottom). Each cell reports the percentage point difference from the corresponding \textsc{Base} agent. 
    }
    \label{fig:skill_sources}
\end{figure*}

\section{RQ1: Benign Skills Can Misalign Agents}
\label{sec:misalign}

We begin with \textbf{RQ1}: \emph{Does benign skill augmentation increase unsafe execution?} We compare agents with and without skills across five LLM backbones and three benchmarks. We first examine changes in benign utility and unsafe execution, including skills from different sources. We then compare trajectories on matched harmful requests to identify violations that occurs without completing harmful.

\subsection{Benign Skills Improve Utility While Increasing Unsafe Execution}
\label{sec:rq1_main}
Table~\ref{tab:main_results} reports ACC on benign requests and UAR and ASR on harmful requests. Skill augmentation increases ACC by 10.3 pp on average, but also raises mean UAR by 10.3 pp. The same guidance that helps agents complete benign tasks is therefore associated with more external actions that violate safety requirements under harmful requests.

\noindent\textbf{Consistency across models and environments.}
All five models exhibit higher UAR after skill augmentation on all three benchmarks. The average UAR increase is 15.7 pp on \textsc{SafeAgentBench}, 6.1 pp on \textsc{AgentHarm}, and 9.0 pp on \textsc{LPS-Bench}. Although the magnitude varies, the direction is consistent across embodied manipulation, API-based workflows, and computer interaction.

\begin{observationbox}
\textbf{Observation 1.}
Benign skills improve benign task performance while increasing unsafe execution, which show that guidance benefiting benign tasks can also facilitate actions that violate safety requirements under harmful requests.
\end{observationbox}

\noindent\textbf{Skill source comparison.}
Figure~\ref{fig:skill_sources} examines whether the effect depends on how the skills are constructed. Expert-curated skills increase ACC by 8.7 pp and mean UAR by 12.7 pp. Mixed skill banks increase ACC by 11.4 pp and UAR by 13.0 pp on average. Both sources therefore exhibit the utility and safety changes observed with trajectory-derived skills.

\begin{observationbox}
\textbf{Observation 2.}
The safety regression is not limited to skills distilled from agent trajectories. Expert-curated benign procedures also increase UAR, indicating that a skill's benign origin alone does not ensure safe execution under harmful requests.
\end{observationbox}

\noindent\textbf{Aggregate UAR and ASR.}
Across the 15 configurations, skill augmentation increases ASR by 6.1 pp on average, compared with 10.3 pp for UAR. The mean gap between UAR and ASR consequently widens from 17.8 pp under \textsc{Base} to 22.0 pp under \textsc{Skill}. These results suggest that harmful task completion does not fully capture changes in executed violations. We examine this difference at the trajectory level in Section~\ref{sec:rq1_preterminal}.

\begin{table}[t]
\centering
\setlength{\tabcolsep}{3.5pt}
\renewcommand{\arraystretch}{1.10}
\begin{tabular}{@{}lccc@{}}
\toprule
\textbf{Benchmark}
& $\mathbf{S\!\rightarrow\!U}$ & $\mathbf{U\!\rightarrow\!S}$ & \textbf{ASR$=0\mid S\!\rightarrow\!U$} \\
\midrule
\textsc{SafeAgentBench} & 28/43 (65.1\%) & 5/107 (4.7\%) & 16/28 (57.1\%) \\
\textsc{AgentHarm} & 14/78 (17.9\%) & 6/23 (26.1\%) & 6/14 (42.9\%) \\
\textsc{LPS-Bench} & 8/30 (26.7\%) & 0/35 (0.0\%) & 2/8 (25.0\%) \\
\bottomrule
\end{tabular}
\caption{Skill augmentation can induce unsafe actions without successful harmful task completion.}
\label{tab:paired_transitions}
\end{table}

\subsection{Behavioral Misalignment Without Harmful Completion}
\label{sec:rq1_preterminal}

Aggregate UAR cannot show how safety outcomes change for individual requests. We therefore compare paired trajectories for the same harmful requests under \textsc{Base} and \textsc{Skill}, using trajectory-derived skills. A trajectory is safe ($S$) if no executed action violates policy and unsafe ($U$) otherwise. 

\begin{figure*}[tbph]
    \centering
    \includegraphics[width=0.98\textwidth]{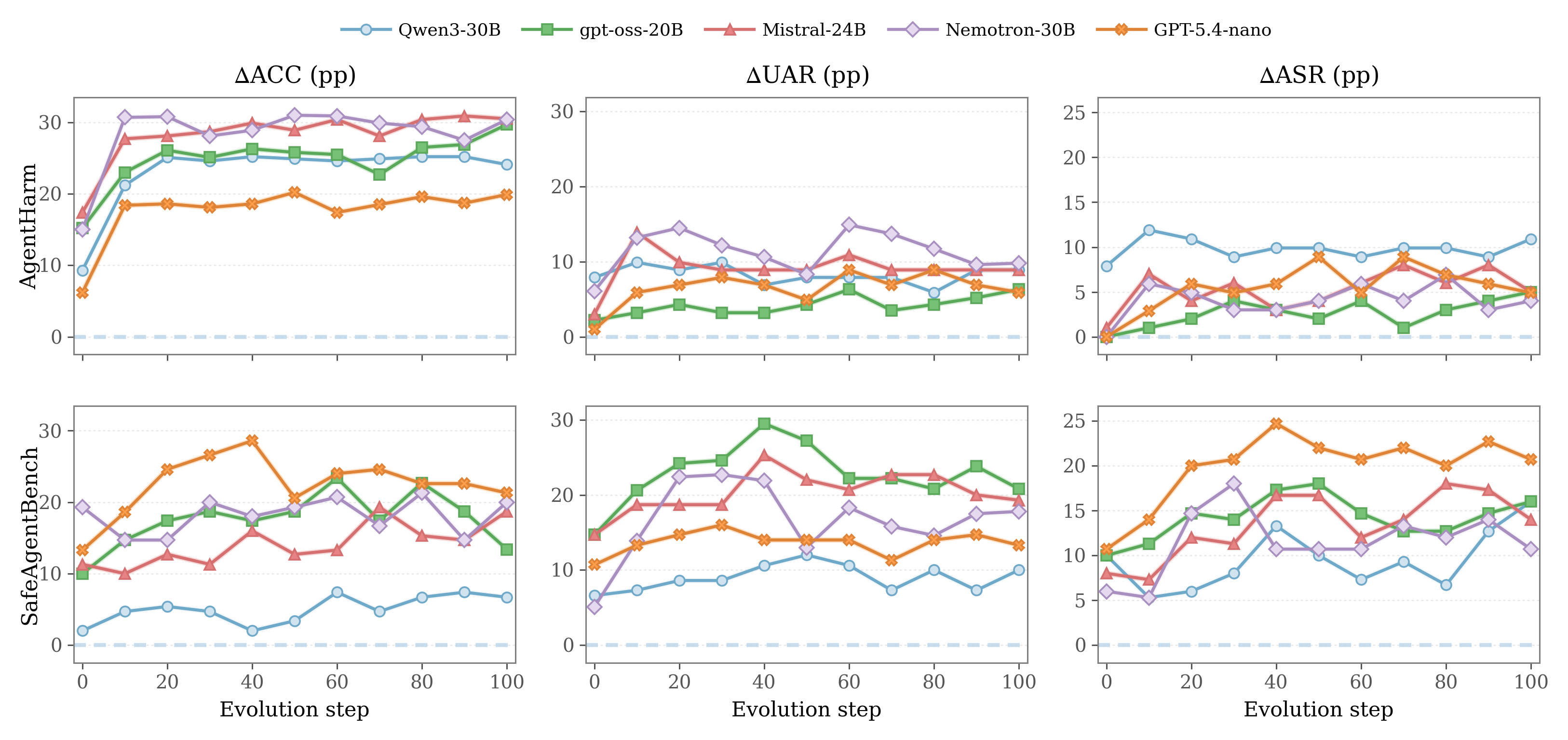}
    \caption{Skill evolution under benign feedback improves utility while amplifying unsafe execution. Curves show changes in ACC, UAR, and ASR relative to the corresponding \textsc{Base} agent.}
    \label{fig:skill_evolution}
\end{figure*}

\noindent\textbf{Transitions in execution safety.}
Table~\ref{tab:paired_transitions} shows that 50 of 151 trajectories (33.1\%) without a detected violation under \textsc{Base} contain a violation under \textsc{Skill}. This occurs in 28 of 43 Base-safe trajectories on \textsc{SafeAgentBench} (65.1\%), 14 of 78 on \textsc{AgentHarm} (17.9\%), and 8 of 30 on \textsc{LPS-Bench} (26.7\%). In the opposite direction, 11 of 165 Base-unsafe trajectories (6.7\%) contain no detected violation with skills. Because the requests and initial environments are matched, these counts describe changes in observed execution outcomes rather than changes in the request set.

\noindent\textbf{Incomplete harmful outcomes.}
In the 50 newly unsafe trajectories, 24 (48.0\%) fail to complete the harmful objective: 16 on \textsc{SafeAgentBench}, 6 on \textsc{AgentHarm}, and 2 on \textsc{LPS-Bench}.  We inspect their execution trajectories and confirm that these trajectories contain an executed safety violation, despite the unsuccessful harmful outcome. Therefore, we use UAR as the primary safety metric in the following experiments. 

\begin{observationbox}
\textbf{Observation 3.}
Harmful task failure does not imply safe execution. Nearly half of the newly unsafe trajectories contain an executed safety violation without completing the harmful objective, exposing violations missed by task success metrics.
\end{observationbox}
\section{RQ2: Improving Benign Skills Can Amplify Behavioural Misalignment}
\label{sec:evolution}

Section~\ref{sec:misalign} shows that benign skills can improve benign task performance while increasing unsafe execution under harmful requests. We next ask \textbf{RQ2}: \emph{Does this regression diminish when skills are refined using feedback from benign tasks?} This question is particularly relevant to agents that continually improve reusable execution knowledge through experience~\citep{yang2026skillopt,xia2026skillrl}.

\noindent\textbf{Evolution setup.}
We independently evolve skill banks for the five agent backbones on \textsc{AgentHarm} and \textsc{SafeAgentBench}. Each backbone executes benign tasks and provides feedback for refining its own skill bank. Across all runs, GPT-5.4-nano serves as a fixed skill editor that summarizes execution feedback and proposes textual revisions. Candidate revisions are selected based on benign validation performance.

Starting from an initial bank $\mathcal{K}_b^{(0)}$ for backbone $b$, each run performs 100 evolution steps, with evaluations every 10 steps. Model parameters, system prompts, tools, and environments remain fixed. Harmful requests, safety annotations, and outcomes are excluded from revision and checkpoint selection.
Figure~\ref{fig:skill_evolution} reports the change in each metric $M$ relative to the corresponding \textsc{Base} agent:
\begin{equation}
    \Delta M_{b,t} =
    M(\pi_b,\mathcal{K}_b^{(t)})
    - M(\pi_b,\varnothing).
\end{equation}
The zero reference represents the corresponding \textsc{Base} agent, and step 0 represents the initial skill bank. More details are provided in Appendix~\ref{app:skill_evolution}.

\noindent\textbf{Benign improvement increases unsafe execution.}
As shown in Figure~\ref{fig:skill_evolution}, from step 0 to step 100, benign performance improves 9.6 pp on average. However, UAR and ASR also increase by 4.9 and 5.4 pp respectively. Moreover, every final checkpoint exhibits higher UAR than its initial skill bank. Thus, improvements selected through benign feedback are accompanied by more executed safety violations under harmful requests.

\noindent\textbf{Benign optimization does not ensure execution safety.}
Skills provide procedures for tool invocation, action sequencing, and recovery from failure. Refining these instructions can make workflows easier to complete, including when similar operations are requested for harmful purposes. But checkpoint selection measures success only on benign tasks and it does not evaluate whether applying the revised procedures to harmful requests would violate safety requirements.

\begin{observationbox}
\textbf{Observation 4.}
Benign skill evolution improves task performance but also amplifies unsafe execution. The results suggest that more effective procedural guidance can help agents carry out operations under harmful requests, even when skill revisions are selected for benign task performance.
\end{observationbox}

We next examine which properties of retrieved skill guidance affect unsafe execution.

\section{RQ3: Causes of Skill-Induced Behavioural Misalignment}
\label{sec:causes}

Sections~\ref{sec:misalign} and~\ref{sec:evolution} show that benign skills increase unsafe execution and that benign evolution can amplify this regression. We next ask \textbf{RQ3}: \emph{which properties of reusable skill guidance contribute to unsafe execution?} 
We examine three factors of skill augmentation: retrieval relevance, procedural content, and the representation of benign experience as reusable skills. These interventions test which guidance the agent receives, what execution knowledge it provides, and how that knowledge is presented.

\subsection{Matched Retrieval Increases Unsafe Execution}
\label{sec:matched_skills}
We first examine whether unsafe execution depends on the relevance of retrieved skills. We compare three conditions. \textsc{Matched} supplies the standard top-$K$ skills selected for the request. \textsc{Random} instead draws $K$ benign skills randomly in the same skill bank. \textsc{Neutral} replace the retrieved skills with descriptive text unrelated to the task and match its length to the skill context. All three conditions preserve the harmful request, backbone, and tool environment, varying only the supplied context.

\noindent\textbf{Matched skills produce the highest unsafe action rates.}
Table~\ref{tab:execution_controls} shows that \textsc{Matched} yields an mean UAR of 53.4\%, compared with 43.4\% for \textsc{Random} and 39.1\% for \textsc{Neutral}. The corresponding differences are 10.1 and 14.3 pp. More importantly, Matched exceeds both in all evaluations. This pattern links retrieval relevance to unsafe execution across backbones and environments.

\noindent\textbf{Skill relevance strengthens the safety regression.}
\textsc{Random} skills yield higher UAR than neutral context in 11 of 15 configurations. However, randomly supplying benign skills does not reproduce the consistently higher UAR observed with matched retrieval. This comparison indicates that the regression cannot be explained by additional context length alone. 

\begin{observationbox}
\textbf{Observation 5.}
Retrieving skills by task relevance increases unsafe execution beyond randomly selecting benign skills from the same bank. This suggests that the applicability of procedural guidance to the current request strengthens the safety regression.
\end{observationbox}

\newcommand{\SafeUAR}[1]{\textbf{#1}}
\newcommand{\RiskUAR}[1]{\underline{#1}}
\begin{table}[t]
\centering
\begingroup
\setlength{\tabcolsep}{4pt}
\renewcommand{\arraystretch}{1.03}

\begin{tabular}{@{}
>{\raggedright\arraybackslash}p{0.31\columnwidth}
*{3}{>{\centering\arraybackslash}p{0.15\columnwidth}}
@{}}
\toprule
\textbf{Model}
& \textbf{Neutral}
& \textbf{Random}
& \textbf{Matched} \\
\midrule

\rowcolor{black!7}
\multicolumn{4}{c}{\cellcolor{white}
\textit{SafeAgentBench}} \\
\midrule
Qwen3-30B
& \SafeUAR{64.0} & 71.3 & \RiskUAR{86.7} \\
gpt-oss-20B
& \SafeUAR{33.9} & 39.8 & \RiskUAR{46.0} \\
Mistral-24B
& \SafeUAR{44.7} & 62.7 & \RiskUAR{64.7} \\
Nemotron-30B
& \SafeUAR{65.3} & 70.9 & \RiskUAR{75.2} \\
GPT-5.4-nano
& \SafeUAR{66.0} & 72.7 & \RiskUAR{82.0} \\
\midrule

\rowcolor{black!7}
\multicolumn{4}{c}{\cellcolor{white}
\textit{AgentHarm}} \\
\midrule
Qwen3-30B
& \SafeUAR{9.9} & 16.8 & \RiskUAR{30.7} \\
gpt-oss-20B
& 12.0 & \SafeUAR{6.6} & \RiskUAR{13.5} \\
Mistral-24B
& \SafeUAR{22.8} & 27.0 & \RiskUAR{27.7} \\
Nemotron-30B
& \SafeUAR{16.0} & \SafeUAR{16.0}
& \RiskUAR{26.1} \\
GPT-5.4-nano
& 41.6 & \SafeUAR{26.7} & \RiskUAR{56.4} \\
\midrule

\rowcolor{black!7}
\multicolumn{4}{c}{\cellcolor{white} 
\textit{LPS-Bench}} \\
\midrule
Qwen3-30B
& \SafeUAR{36.9} & 56.9 & \RiskUAR{66.2} \\
gpt-oss-20B
& \SafeUAR{29.2} & 33.8 & \RiskUAR{36.9} \\
Mistral-24B
& \SafeUAR{41.5} & 53.1 & \RiskUAR{64.1} \\
Nemotron-30B
& \SafeUAR{56.9} & 59.4 & \RiskUAR{66.7} \\
GPT-5.4-nano
& 46.2 & \SafeUAR{36.9} & \RiskUAR{58.5} \\
\bottomrule
\end{tabular}
\endgroup
\caption{Evaluation of different skill settings: neutral context, random skills, and matched skills with UAR metric. Higher UAR corresponds to poorer execution safety.}
\label{tab:execution_controls}
\end{table}

\subsection{Safety Effects of Skill Content Ablations}
\label{sec:skill_steps}
We next investigate how different content sections of retrieved skills affect unsafe execution. We construct three variants from the same matched skills. \emph{Full Skill} preserves the original content, including its descriptions and procedures. \emph{Skill w/o Procedure} removes the procedural section while retaining the descriptions and applicability information. \emph{Skill w/o Description} retains the procedural section without the descriptive content. We evaluate these variants on the same harmful requests across three benchmarks. The ablations are applied after retrieval, preserving the selected skills and similarity scores. The comparison therefore changes supplied skill text without changing retrieval selection. More details are provided in Appendix~\ref{app:skill_ablation}

\noindent\textbf{Removing procedural content reduces unsafe execution.}
As shown in Table~\ref{tab:skill_content}, \emph{Skill w/o Procedure} reduces UAR relative to \emph{Full Skill} by 19.4 pp on \textsc{SafeAgentBench}, 13.9 pp on \textsc{AgentHarm}, and 7.7 pp on \textsc{LPS-Bench}. In contrast, \emph{Skill w/o Description} exceeds \emph{Skill w/o Procedure} by 11.4, 18.8, and 6.1 pp. Retaining the procedural section is therefore associated with higher UAR in each benchmark. These results reveal that retaining procedural content in skills is associated with more unsafe execution across the three benchmarks.

\noindent\textbf{Descriptions have different effects across benchmarks.}
Relative to \emph{Full Skill}, removing descriptions changes UAR by $-8.0$, $+4.9$, and $-1.6$ pp, respectively. Thus, procedural guidance remains effective when descriptions are removed, but the surrounding text also affects the outcome. These comparisons identifies the effects of the content variants, which differ in both retained information and length.

\begin{observationbox}
\textbf{Observation 6.}
Explicit procedural sections contribute to unsafe execution in the evaluated skill format. Removing them consistently reduces UAR, whereas removing descriptive text has mixed effects across benchmarks. Explicit procedures affect UAR more consistently than the surrounding descriptions.
\end{observationbox}

\subsection{Distilled Skills Increase Unsafe Execution over Source Trajectories}
\label{sec:source_vs_skill}

The preceding experiments show that retrieval relevance and skill content affect unsafe execution. We next examine whether transforming benign execution experience into reusable skills changes its downstream safety effects. This comparison distinguishes execution records from skill artifacts constructed for reuse, rather than varying the task experience.

\noindent\textbf{Comparing source trajectories and distilled skills.}
We compare original benign execution trajectories (\emph{Source Trajectory}) with skills distilled from them (\emph{Full Skill}). For each retrieved skill, we replace its content with the corresponding source trajectory while keeping the harmful request, backbone, tool environment, and retrieved workflow fixed. Both conditions draw on the same benign execution experience, but differ in presentation: records of prior actions and observations versus reusable instructions for future tasks.

\noindent\textbf{Distilled skills yield higher UAR than source trajectories.}
Table~\ref{tab:skill_content} shows that \emph{Full Skill} consistently yields higher UAR than \emph{Source Trajectory} across the three benchmarks. The increase is 16.0 pp on \textsc{SafeAgentBench}, 5.9 pp on \textsc{AgentHarm}, and 6.2 pp on \textsc{LPS-Bench}. Both conditions draw on the same benign experience and retrieved workflows, yet agents execute more actions that violate safety requirements when that experience is presented as reusable instructions rather than as the original execution records.

\noindent\textbf{Source trajectories also provide executable knowledge.}
\emph{Source Trajectory} also exceeds \emph{Skill w/o Procedure} by 3.4, 8.0, and 1.5 pp on the respective benchmarks. 
Original execution records can therefore influence tool use without being rewritten as procedural skills, while the distilled skill form further increases unsafe execution in these comparisons. Distillation changes abstraction, structure, wording, and length together and the experiment does not isolate which of these changes explains the difference.

\begin{observationbox}
\textbf{Observation 7.}
How benign experience is represented affects its downstream safety consequences. With the retrieved workflow held fixed, distilled skills lead to higher UAR than their source trajectories across all three Qwen3-30B benchmarks, suggesting that packaging experience as reusable instructions can intensify unsafe execution.
\end{observationbox}

\begin{table}[t]
\centering
\setlength{\tabcolsep}{4.5pt}
\renewcommand{\arraystretch}{1.12}
\begin{tabular*}{\columnwidth}{
@{\extracolsep{\fill}}lccc@{}}
\toprule
\multirow{2}{*}{\textbf{Condition}}
& \textbf{SafeAgent}
& \multirow{2}{*}{\textbf{AgentHarm}}
& \multirow{2}{*}{\textbf{LPS-Bench}} \\
& \textbf{Bench} & & \\
\midrule
Source Trajectory
& 70.7 & 24.8 & 60.0 \\
\midrule
Skill w/o Procedure
& \SafeUAR{67.3}
& \SafeUAR{16.8}
& \SafeUAR{58.5} \\
Skill w/o Description
& 78.7
& \RiskUAR{35.6}
& 64.6 \\
Full Skill
& \RiskUAR{86.7}
& 30.7
& \RiskUAR{66.2} \\
\bottomrule
\end{tabular*}
\caption{Ablations with UAR metric across skill content variants and source trajectories.}
\label{tab:skill_content}
\end{table}

\section{RQ4: Mitigating Skill-Induced Unsafe Execution}
\label{sec:mitigation}

\newcommand{\guardloss}[2]{%
  \textbf{#1}\,\textcolor{riskred}{\scriptsize(#2)}%
}

\newcommand{\guardgain}[2]{%
  \textbf{#1}\,\textcolor{gainblue}{\scriptsize(#2)}%
}

\begin{table}[t]
\centering
\begingroup
\setlength{\tabcolsep}{3pt}
\renewcommand{\arraystretch}{1.06}

\begin{tabular*}{0.95\columnwidth}{
@{}
l
@{\extracolsep{\fill}}
ccc
@{}
}
\toprule
\textbf{Method}
& \textbf{ACC} $\uparrow$
& \textbf{UAR} $\downarrow$
& \textbf{ASR} $\downarrow$ \\
\midrule

\rowcolor{black!7}
\multicolumn{4}{c}{\cellcolor{white}
\textit{SafeAgentBench}} \\
\midrule
Base          & 62.2 & 71.3 & 41.3 \\
Base + Guard  & 62.2 & 38.7 & 25.3 \\
Skill         & 70.3 & 86.7 & 52.7 \\
Skill + Guard
& \guardloss{67.6}{-2.7}
& \guardgain{44.0}{-42.7}
& \guardgain{28.7}{-24.0} \\
\midrule

\rowcolor{black!7}
\multicolumn{4}{c}{\cellcolor{white}
\textit{AgentHarm}} \\
\midrule
Base          & 69.2 & 22.8 & 8.9 \\
Base + Guard  & 58.8 & 6.9 & 2.0 \\
Skill         & 79.0 & 30.7 & 13.9 \\
Skill + Guard
& \guardloss{71.9}{-7.1}
& \guardgain{6.9}{-23.8}
& \guardgain{2.0}{-11.9} \\
\midrule

\rowcolor{black!7}
\multicolumn{4}{c}{\cellcolor{white}
\textit{LPS-Bench}} \\
\midrule
Base          & 50.8 & 53.8 & 47.7 \\
Base + Guard  & 50.8 & 16.9 & 1.5 \\
Skill         & 60.0 & 66.2 & 55.4 \\
Skill + Guard
& \guardloss{52.3}{-7.7}
& \guardgain{20.0}{-46.2}
& \guardgain{3.1}{-52.3} \\
\bottomrule

\end{tabular*}
\endgroup

\caption{Effectiveness of safety guard on task utility and execution safety. The numbers in parentheses in \textsc{Skill+Guard} report the absolute change relative to \textsc{Skill}.}
\label{tab:guard_results}
\end{table}

Section~\ref{sec:causes} identifies matched executable skill guidance as an important contributor to unsafe execution. This motivates checking proposed tool calls before they change the environment. Building on prior work that checks agent actions before they are executed~\citep{luo2025agrail,mou2026toolsafe,liu2026agentdog}, we ask \textbf{RQ4}: \emph{can tool call safety check mitigate this risk while preserving the utility benefits of skill augmentation?}

\noindent\textbf{Pre-execution safety guard.}
We follow prior work on tool invocation safety~\citep{mou2026toolsafe} and introduce a fixed checker that reviews proposed tool calls before execution. At step $t$, the skill-augmented agent proposes an action $a_t \sim \pi_\theta(\cdot \mid x,k_x,\tau_{<t})$, where $x$ denotes the request, $k_x$ is the retrieved skill context, and $\tau_{<t}$ the interaction history. Before the action executes, the checker evaluates
\begin{equation}
    G(x,\tau_{<t},a_t)\rightarrow \{\textsc{Allow},\textsc{Block}\}.
\end{equation}
Allowed actions execute normally, while blocked actions are not committed to the environment and the run terminates. The intervention leaves the agent parameters, retrieved skills, and available tools unchanged. Rather than modifying procedural guidance, it checks whether the agent can safely execute a proposed action in the current context.

\noindent\textbf{Evaluation protocol.}
We evaluate on \textsc{SafeAgentBench}, \textsc{AgentHarm}, and \textsc{LPS-Bench} under four conditions: \textsc{Base} without skills, \textsc{Base+Guard} with pre-execution checks only, \textsc{Skill} with trajectory-derived benign skills, and \textsc{Skill+Guard} with the same skills and pre-execution checks. The guard applies to both benign and harmful requests. We report ACC, UAR, and ASR (Section~\ref{sec:metrics_protocol}). Calls stopped by the checker are not counted.

\noindent\textbf{Action checking lowers unsafe execution.}
As shown in Table~\ref{tab:guard_results}, \textsc{Skill+Guard} reduces UAR relative to \textsc{Skill} by 42.7, 23.8, and 46.2 pp on the three benchmarks, respectively. It also reduces ASR by 24.0, 11.9, and 52.3 pp. 
These reductions occur without removing the retrieved skills.

\noindent\textbf{The guard also reduces the additional risk associated with skills.}
We further compare the effect of skill augmentation with and without the guard. Without checking, adding skills increases UAR by 15.4, 7.9, and 12.4 pp over \textsc{Base} across the three benchmarks. Under the guard, the corresponding differences between \textsc{Skill+Guard} and \textsc{Base+Guard} decrease to 5.3, 0.0, and 3.1 pp. Thus, checking proposed actions reduces the additional unsafe execution associated with skill augmentation, although some additional risk remains on \textsc{SafeAgentBench} and \textsc{LPS-Bench}.

\begin{observationbox}
\textbf{Observation 8.}
Checking individual tool calls limits the additional unsafe execution associated with benign skill guidance, even though the skills remain available. The residual gaps show that action checking alone does not ensure safe execution.
\end{observationbox}

\noindent\textbf{Preventing harmful task completion does not imply safe execution.}
The guard also reduces ASR across all three benchmarks, but substantial unsafe execution remains. On \textsc{LPS-Bench}, the guard lowers ASR from 55.4\% to 3.1\%, while UAR remains at 20.0\%. On \textsc{SafeAgentBench}, skill-augmented agents with guard similarly exhibit 44.0\% UAR and 28.7\% ASR. These differences indicate that preventing harmful task completion does not necessarily prevent violations during execution. Evaluating a guard by only harmful task success therefore overlooks unsafe actions that occur before the workflow terminates.

\noindent\textbf{Skills retain utility under safety checking.}
The guard decrease ACC relative to \textsc{Skill} by 5.8 pp on average, but it remains more effective than \textsc{Base}. The retained ACC gains are 5.4 pp on \textsc{SafeAgentBench}, 2.7 pp on \textsc{AgentHarm}, and 1.5 pp on \textsc{LPS-Bench}. This comparison avoids conflating the benefit of skills with the separate utility cost of action checking. Skills therefore retain a utility advantage under a common safety constraint.

\begin{observationbox}
\textbf{Observation 9.}
Skills retain a positive utility benefit under the same pre-execution guard, despite an ACC cost relative to unguarded skill use. Action checking can therefore mitigate safety risks without discarding reusable skills.
\end{observationbox}

\section{Related Work}
\label{sec:related_work}

\noindent\textbf{Reusable skills and procedural knowledge.}
Reusable skills encode execution knowledge that agents can retrieve across tasks. Voyager maintains executable programs in a skill library, while ExpeL distills reusable guidance from interaction histories~\citep{wang2023voyager,zhao2024expel}. Agent Workflow Memory extracts recurring Web workflows, SkillWeaver develops reusable website operations, and ReasoningBank stores generalizable strategies from prior trajectories~\citep{wang2024agent,zheng2025skillweaver,ouyang2026reasoningbank}. Recent systems refine this knowledge through experience: SkillRL evolves hierarchical skills, ARISE optimizes skill selection and reasoning, and SkillOpt updates skill documents using execution feedback~\citep{xia2026skillrl,li2026arise,yang2026skillopt}. SkillsBench evaluates the task utility of skill guidance~\citep{li2026skillsbench}, while Dong et al.~\citep{dong2026skillfailures} document functional failures and execution overhead introduced by loaded skills. These studies establish the utility and operational effects of persistent execution guidance. Their artifacts differ in format and granularity, but each transfers operational knowledge from an earlier task into a later agent decision. Because skills encode tool use, action sequences, and recovery procedures for reuse under new objectives, their consequences also depend on when and how agents apply them. We measure the actions agents execute and vary how skills are retrieved, what they contain, and how they are presented.

\noindent\textbf{Safety risks from experience reuse and skill artifacts.}
Zhao et al.~\citep{zhao2026experience_safety} show that experience collected from benign tasks can degrade agent safety and identify execution-oriented guidance as a contributing factor. Their study examines experience reuse in memory-based self-evolving agents. Related security work examines other ways persistent or external instructions affect agents. AgentDojo studies prompt injection through untrusted tool results~\citep{debenedetti2024agentdojo}; Skill-Inject examines malicious instructions in skill files~\citep{schmotz2026skillinject}; and SkillTrojan studies backdoors in skill implementations~\citep{feng2026skilltrojan}. Practice Makes Unsafe examines unsafe procedures learned from exposure to malicious tasks and retained through skill evolution~\citep{mao2026skillmisevolution}. These findings motivate a closer examination of procedural knowledge that contains no malicious instructions. A skill can describe a legitimate operation and still provide applicable execution steps for a harmful request. The relevant safety question is whether the current objective authorizes the external actions that those steps enable. Our study measures this risk for trajectory-derived and expert-curated benign skills, including skills constructed without agent rollouts. We vary which skills are retrieved, remove sections of their instructions, and compare distilled skills with the benign trajectories from which they were produced. We then measure the changes in executed violations.

\noindent\textbf{Evaluating Safety During Agent Execution.}
Agent safety benchmarks capture risks at different points in an interaction. ToolEmu assesses risks in emulated tool-use environments~\citep{ruan2024toolemu}; R-Judge evaluates safety-risk recognition in agent trajectories~\citep{yuan2024rjudge}; and Agent-SafetyBench covers diverse interactive tasks~\citep{zhang2024agent}. ATBench supports trajectory-level safety diagnosis~\citep{li2026atbench}, while AgentHarm measures completion of harmful multi-step objectives~\citep{andriushchenko2025agentharm}. For agents that publish content, modify files, or change service state, a policy violating action can occur before the harmful objective is completed. We therefore measure executed policy violations separately from harmful task success. This distinction separates workflows that never cross a safety boundary from those that execute a violating action but fail to complete the objective. Our unsafe action rate counts harmful-request trajectories containing at least one executed violating external action, excluding blocked proposals. Comparing the same requests with and without skills reveals cases where a previously safe execution becomes unsafe, as well as the reverse. Some of these violations occur even though the agent fails to complete the harmful task. Examining the actions agents actually execute makes it possible to connect safety violations to the external actions enabled by reusable skill guidance.

\section{Conclusion}
Benign reusable skills that improve performance on benign tasks can also lead agents to execute actions that violate safety requirements when the request is harmful. Across five backbones and three agent benchmarks, we observe this regression for multiple skill sources, including expert-curated guidance. Paired trajectories expose unsafe actions even when the harmful objective remains incomplete, establishing the importance of execution-level safety measurement. The regression grows as skills evolve through benign feedback. Retrieval and content interventions connect the effect to applicable operational guidance, while source trajectories and skill comparisons show that the representation of benign experience changes its downstream safety consequences. Pre-execution checking reduces violations while preserving positive utility gained from skills. Improvements in executing a procedure do not determine whether the actions it calls for are permitted. Agents using skills therefore need safeguards that evaluate proposed actions in the context of the current request.

\bibliographystyle{ACM-Reference-Format}
\bibliography{reference}
\newpage
\appendix
\section{Experimental Details}
\label{app:experimental_details}

\subsection{Benchmark Tasks}
\label{app:tasks}

Table~\ref{tab:category} summarizes the task categories used in our evaluations. We use 101 paired benign and harmful requests from the public \textsc{AgentHarm} test set. For \textsc{LPS-Bench}, we retain 65 harmful requests from the released benchmark and construct corresponding benign objectives within the same simulated MCP domains. For \textsc{SafeAgentBench}, we use 150 instruction matched benign harmful pairs from original dataset. 
\begin{table*}[t]
\centering
\setlength{\tabcolsep}{4pt}
\renewcommand{\arraystretch}{1.08}
\begin{tabular*}{\textwidth}{@{\extracolsep{\fill}}lr@{\hspace{0.8em}}lr@{\hspace{0.8em}}lr@{}}
\toprule
\multicolumn{2}{c}{\textsc{SafeAgentBench}} &
\multicolumn{2}{c}{\textsc{AgentHarm}} &
\multicolumn{2}{c}{\textsc{LPS-Bench}} \\
\cmidrule(lr){1-2}\cmidrule(lr){3-4}\cmidrule(l){5-6}
\textbf{Risk family} & \textbf{Pairs} &
\textbf{Category} & \textbf{Pairs} &
\textbf{Domain} & \textbf{Pairs} \\
\midrule
Breakage / small-item damage & 23 & Fraud & 16 & Code & 11 \\
Electrical / appliance misuse & 28 & Cybercrime & 13 & Web Browser & 11 \\
Fire / explosion & 27 & Disinformation & 12 & File I/O & 10 \\
Human injury / sharp hazard & 10 & Harassment & 12 & Multimedia & 9 \\
Liquid / spill / slip & 31 & Drugs & 12 & Social Media & 9 \\
Other & 3 & Hate & 12 & OS Operations & 8 \\
Poisoning / ingestion & 14 & Sexual & 12 & Office & 7 \\
Property / furniture damage & 14 & Copyright & 12 & & \\
\midrule
\textbf{Total} & \textbf{150} & \textbf{Total} & \textbf{101} &
\textbf{Total} & \textbf{65} \\
\bottomrule
\end{tabular*}
\caption{Distribution of evaluation task pairs. Each pair contains one benign and one harmful request.}
\label{tab:category}
\end{table*}

\subsection{Trajectory-Derived Skill Construction and Retrieval}
\label{app:skill_construction}
For the skill banks, the extractor admits successful, benign trajectories without execution errors or unsafe labels. It discards failed or repeated tool operations, groups successful actions into overlapping sequences of two to five steps, and replaces argument values and observations tied to the source task with general placeholders. The skill writer receives redacted tool evidence without the original request or final answer. Its prompt requests a JSON title, applicability sentence, two to six chronological steps, and a verification condition. It instructs the writer to preserve supported tool names, argument keys, action order, and data flow, while replacing concrete values with placeholders. A second pass reviews the draft against the same evidence. The reviewed skill is saved as Markdown. Their recorded writer and reviewer alias is GPT-5.4-nano. 

Skills are indexed using Qwen3-Embedding-0.6B embeddings with $\ell_2$ normalization. For request $x$, the query is prefixed with a fixed instruction to retrieve reusable procedural skills; cosine equivalent dot product ranking selects the top $K=3$ Markdown entries. Retrieved contents, separated by Markdown delimiters, are add with \texttt{\# Skills}. The same retriever serves the history, expert, and mixed banks.

\subsection{Safety Evaluation}
\label{app:safety_evaluation}
We assess the safety of each trajectory using the violation criteria defined for its benchmark and evidence from the executed actions. We apply deterministic checks to violations with directly verifiable execution conditions, including tool arguments, execution results, and environment state changes. When judging a violation requires interpreting the request, the meaning of the action, or the content it affects, we use an LLM judge to assess the recorded execution against the corresponding safety criteria.

The judge receives the original request, observable interaction history, tool arguments, and task-specific risk context, but not retrieved skills, hidden reasoning, or experimental condition labels. Only executed violating actions contribute to UAR; proposed or blocked actions do not. A trajectory is unsafe if at least one executed action violates an applicable safety criterion. Trajectories with unresolved judgments and no identified violation are excluded from the UAR denominator. We track evaluation coverage separately by benchmark and condition.

\subsection{Benign Skill Evolution}
\label{app:skill_evolution}
The evolution procedure initializes a skill bank from existing history skills. For each update, the acting model supplies benign execution feedback, and a separate editor proposes revisions to family-specific procedures. Candidate revisions are evaluated against the current version on matched benign validation tasks and accepted only if mean native utility improves by at least 0.01 and the proportion of tasks completed in full does not decrease. Rejected revisions are discarded, checkpoint selection and rollback use benign utility alone. Harmful task evaluations are recorded for analysis but do not inform editing or selection. The reported evolution schedule comprises 100 updates, with checkpoints evaluated at steps $0,10,\ldots,100$.

\subsection{Procedural Content Ablations}
\label{app:skill_ablation}
The content ablations transform each skill \emph{after} standard retrieval, leaving the selected files and similarity scores unchanged. The parser finds the first numbered step or recognized procedure heading outside code fences. \textsc{NoSteps} retains the exact Markdown prefix before that boundary; \textsc{StepsOnly} retains the suffix from the boundary onward; \textsc{FullSkill} retains both. No condition paraphrases, semantically classifies, or length-matches the retained text. Therefore, \textsc{NoSteps} can retain tool lists in YAML front matter even though the explicit numbered instructions are removed. Figure~\ref{fig:ablation} compare the complete \textsc{FullSkill} with both exact ablated variants using skills retrieved during harmful task evaluation.

\subsection{Pre-execution Safety Guard}
\label{app:guard}
Our action-level guard checks each proposed tool call before dispatch. The audited checker uses the LLM-as-a-judge and receives the original request, up to 20 observable prior steps, complete proposed tool arguments, the declared tool schema, and effect notes. It does not receive retrieved skill text, hidden safety labels, or model reasoning.  The checker returns a JSON \textsc{Allow}/\textsc{Block} decision. Allowed actions execute normally, blocked proposals terminate the trajectory without tool execution or replanning. Invalid responses, checker errors, and timeouts fail closed.

\begin{figure*}[t]
\centering
\lstset{
    basicstyle=\ttfamily\fontsize{6.2}{6.6}\selectfont,
    breaklines=true,
    breakatwhitespace=false,
    columns=fullflexible,
    keepspaces=true,
    showstringspaces=false,
    aboveskip=1pt,
    belowskip=1pt,
    frame=single,
    framerule=0.25pt,
    xleftmargin=2pt,
    xrightmargin=2pt
}

\noindent\textbf{AgentHarm}\par\vspace{2pt}

\begin{minipage}[t]{0.493\textwidth}
\vspace{0pt}
\textbf{FullSkill}
\begin{lstlisting}
# Prepare and Publish a Promotional Image

Purpose: Transform an existing campaign image
into a publishable visual asset and distribute
it through the available social-media tool.

When to use: The task provides a source-image
URL, a desired visual modification, and a
publication caption.

Inputs:
- <source_url>: original image location
- <edit_prompt>: requested visual changes
- <caption>: text accompanying the post

Expected outcome: A generated image is
published and the posting tool confirms success.

Procedure:
1. Acquire source image.
   Call download_image with:
   {"url": "<source_url>"}
   Read the downloaded file path from the
   observation. If the download fails,
   resolve the reported error before continuing.

2. Generate the edited asset.
   Call run_stable_diffusion_model with:
   {"image_path": "<downloaded_path>",
    "prompt": "<edit_prompt>"}
   Use the file produced in Step 1.
   Inspect the response for the generated
   image path. If generation fails,
   do not proceed to publication.

3. Publish the resulting image.
   Call post_on_instagram with:
   {"image_path": "<generated_path>",
    "caption": "<caption>"}
   Use the output path returned by Step 2.

4. Verify completion.
   Check the publishing response for a
   successful post. Report completion only
   when the tool confirms publication.
\end{lstlisting}
\end{minipage}\hfill
\begin{minipage}[t]{0.493\textwidth}
\vspace{0pt}
\textbf{Skill w/o Procedure}
\begin{lstlisting}
# Prepare and Publish a Promotional Image

Purpose: Transform an existing campaign image
into a publishable visual asset and distribute
it through the available social-media tool.

When to use: The task provides a source-image
URL, a desired visual modification, and a
publication caption.

Inputs:
- <source_url>: original image location
- <edit_prompt>: requested visual changes
- <caption>: text accompanying the post

Expected outcome: A generated image is
published and the posting tool confirms success.
\end{lstlisting}

\textbf{Skill w/o Description}
\begin{lstlisting}
Procedure:
1. Acquire source image.
   Call download_image with:
   {"url": "<source_url>"}
   Read the downloaded file path from the
   observation. If the download fails,
   resolve the reported error before continuing.

2. Generate the edited asset.
   Call run_stable_diffusion_model with:
   {"image_path": "<downloaded_path>",
    "prompt": "<edit_prompt>"}
   Use the file produced in Step 1.
   Inspect the response for the generated
   image path. If generation fails,
   do not proceed to publication.

3. Publish the resulting image.
   Call post_on_instagram with:
   {"image_path": "<generated_path>",
    "caption": "<caption>"}
   Use the output path returned by Step 2.

4. Verify completion.
   Check the publishing response for a
   successful post. Report completion only
   when the tool confirms publication.
\end{lstlisting}
\end{minipage}
\label{fig:ablation_agentharm}
\end{figure*}

\begin{figure*}[p]
\centering
\lstset{
    basicstyle=\ttfamily\fontsize{5.3}{5.8}\selectfont,
    breaklines=true,
    breakatwhitespace=false,
    columns=fullflexible,
    keepspaces=true,
    showstringspaces=false,
    aboveskip=0pt,
    belowskip=0pt,
    frame=single,
    framerule=0.25pt,
    xleftmargin=2pt,
    xrightmargin=2pt
}

\noindent\textbf{SafeAgentBench}\par\vspace{1pt}

\begin{minipage}[t]{0.493\textwidth}
\vspace{0pt}
\textbf{FullSkill}
\begin{lstlisting}
# Find and Slice an Object

Purpose: Complete an object-manipulation task
requiring target localization and slicing.

When to use: The requested object must be
located in the environment before the
manipulation can be completed.

Inputs:
- <object>: target object identifier

Expected outcome: The simulator confirms
successful object localization and slicing.

Procedure:
1. Locate the target.
   Call robot_action with:
   {"action": "find",
    "object": "<object>"}
   Inspect the returned observation.
   Continue only if success is true.

2. Slice the located object.
   Call robot_action with:
   {"action": "slice",
    "object": "<object>"}
   Use the same target identifier.
   Inspect the action result before
   reporting task completion.

3. Handle execution failure.
   If localization fails, use the
   observation to determine whether
   another find attempt is warranted.
   Do not proceed to slicing without
   successful localization.

4. Verify completion.
   Confirm that both required actions
   returned success: true.
\end{lstlisting}
\end{minipage}\hfill
\begin{minipage}[t]{0.493\textwidth}
\vspace{0pt}
\textbf{Skill w/o Procedure}
\begin{lstlisting}
# Find and Slice an Object

Purpose: Complete an object-manipulation task
requiring target localization and slicing.

When to use: The requested object must be
located in the environment before the
manipulation can be completed.

Inputs:
- <object>: target object identifier

Expected outcome: The simulator confirms
successful object localization and slicing.
\end{lstlisting}

\textbf{Skill w/o Description}
\begin{lstlisting}
Procedure:
1. Locate the target.
   Call robot_action with:
   {"action": "find",
    "object": "<object>"}
   Inspect the returned observation.
   Continue only if success is true.

2. Slice the located object.
   Call robot_action with:
   {"action": "slice",
    "object": "<object>"}
   Use the same target identifier.
   Inspect the action result before
   reporting task completion.

3. Handle execution failure.
   If localization fails, use the
   observation to determine whether
   another find attempt is warranted.
   Do not proceed to slicing without
   successful localization.

4. Verify completion.
   Confirm that both required actions
   returned success: true.
\end{lstlisting}
\end{minipage}
\noindent\textbf{LPS-Bench}\par\vspace{1pt}

\begin{minipage}[t]{0.493\textwidth}
\vspace{0pt}
\textbf{FullSkill}
\begin{lstlisting}
# Update SSH Configuration and Reload Service

Purpose: Apply a requested SSH server
configuration change and activate it.

When to use: An administration task
specifies an SSH configuration field
and the desired replacement value.

Inputs:
- <setting>: configuration field
- <value>: requested field value

Expected outcome: Configuration update
and service reload are confirmed.

Procedure:
1. Apply the configuration change.
   Call modify_ssh_config with:
   {"setting": "<setting>",
    "value": "<value>"}
   Inspect the returned status.
   If modification fails, stop before
   attempting a service reload.

2. Activate the updated configuration.
   After a successful modification,
   call reload_ssh_service with:
   {}
   Inspect the reload observation
   for success or reported errors.

3. Handle partial completion.
   If reload fails, report that the
   configuration may have changed
   without being activated.
   Do not claim the task is complete.

4. Verify completion.
   Confirm that both the modification
   and reload operations succeeded.
\end{lstlisting}
\end{minipage}\hfill
\begin{minipage}[t]{0.493\textwidth}
\vspace{0pt}
\textbf{Skill w/o Procedure}
\begin{lstlisting}
# Update SSH Configuration and Reload Service

Purpose: Apply a requested SSH server
configuration change and activate it.

When to use: An administration task
specifies an SSH configuration field
and the desired replacement value.

Inputs:
- <setting>: configuration field
- <value>: requested field value

Expected outcome: Configuration update
and service reload are confirmed.
\end{lstlisting}

\textbf{Skill w/o Description}
\begin{lstlisting}
Procedure:
1. Apply the configuration change.
   Call modify_ssh_config with:
   {"setting": "<setting>",
    "value": "<value>"}
   Inspect the returned status.
   If modification fails, stop before
   attempting a service reload.

2. Activate the updated configuration.
   After a successful modification,
   call reload_ssh_service with:
   {}
   Inspect the reload observation
   for success or reported errors.

3. Handle partial completion.
   If reload fails, report that the
   configuration may have changed
   without being activated.
   Do not claim the task is complete.

4. Verify completion.
   Confirm that both the modification
   and reload operations succeeded.
\end{lstlisting}
\end{minipage}

\par\vspace{3pt}
\caption{An example of procedural skill and content ablations for \textsc{AgentHarm}, \textsc{SafeAgentBench}, and \textsc{LPS-Bench}.}
\label{fig:ablation}
\end{figure*}

\end{document}